\documentclass[journal]{IEEEtran}
\usepackage{graphicx}

\usepackage{amsmath}
\usepackage{multirow}
\usepackage{algorithm}
\usepackage{algorithmicx}
\usepackage{algpseudocode}
\usepackage{lipsum}
\usepackage{multicol}
\usepackage{float}
\usepackage{color}
\usepackage{url}
\usepackage{rotating}
\usepackage{caption}
\usepackage{subcaption}

\usepackage{graphics}
\usepackage{epsfig}

\usepackage{epstopdf}

\newcommand{\tabincell}[2]{\begin{tabular}{@{}#1@{}}#2\end{tabular}}

\ifCLASSINFOpdf
\else
\fi
\begin{document}
%
\title{Similarity Grouping-Guided Neural Network Modeling for Maritime Time Series Prediction}
%
%

\author{Yan Li,
        Ryan Wen Liu,
        Zhao Liu,
        and~Jingxian Liu
\thanks{Y. Li, R. W. Liu, Z. Liu, and J. Liu are with the Hubei Key Laboratory of Inland Shipping Technology, School of Navigation, Wuhan University of Technology, Wuhan 430063, China (e-mail: li\_yan@whut.edu.cn; wenliu@whut.edu.cn; zhaoliu@whut.edu.cn; liujingxian@whut.edu.cn).}
\thanks{}
}

%
%

\markboth{Journal of \LaTeX\ Class Files,~Vol.~14, No.~8, August~2015}%
{Shell \MakeLowercase{\textit{et al.}}: Bare Demo of IEEEtran.cls for IEEE Journals}
%



\maketitle

\begin{abstract}
   Reliable and accurate prediction of time series plays a crucial role in maritime industry, such as economic investment, transportation planning, port planning and design, etc. The dynamic growth of maritime time series has the predominantly complex, nonlinear and non-stationary properties. To guarantee high-quality prediction performance, we propose to first adopt the empirical mode decomposition (EMD) and ensemble EMD (EEMD) methods to decompose the original time series into high- and low-frequency components. The low-frequency components can be easily predicted directly through traditional neural network (NN) methods. It is more difficult to predict high-frequency components due to their properties of weak mathematical regularity. To take advantage of the inherent self-similarities within high-frequency components, these components will be divided into several continuous small (overlapping) segments. The grouped segments with high similarities are then selected to form more proper training datasets for traditional NN methods. This regrouping strategy can assist in enhancing the prediction accuracy of high-frequency components. The final prediction result is obtained by integrating the predicted high- and low-frequency components. Our proposed three-step prediction frameworks benefit from the time series decomposition and similar segments grouping. Experiments on both port cargo throughput and vessel traffic flow have illustrated its superior performance in terms of prediction accuracy and robustness.
\end{abstract}
%
%
%
\IEEEpeerreviewmaketitle
\section{Introduction}
\subsection{Background and Related Work}
\IEEEPARstart{D}{ynamic} growth of maritime time series (e.g., port cargo throughput and vessel traffic flow) sometimes has the predominantly complex, nonlinear and non-stationary properties. Their reliable and accurate predictions play an important role in maritime industry, such as economic investment, transportation planning, port planning and design, etc. In the literature \cite{WangTII2018,LvITS2015,ChandraNNLS2015,LamichTIE2017}, many efforts have been recently devoted to effectively predict different types of time series. However, it is still generally difficult to model and predict such non-stationary time series using traditional mathematical methods, such as fuzzy theory \cite{SunNeuro2015,LiuIEEETFS2018}, Kalman filtering \cite{MetiaIEEE2016}, Bayesian models \cite{HuTIE2016}, hybrid framework \cite{JiangEnergy2017}, autoregressive integrated moving average (ARIMA) \cite{KumarVanajakshi2015} and their extensions.

From a machine learning point of view, several techniques have been introduced to the task of time series prediction. The two most-used approaches are certainly conventional neural network (NN) \cite{ZhangIJF1998,ChenTII2017} and latest deep learning (DL) \cite{SchmidhuberNN2015,KhodayarTII2017}. They are both supervised learning methods that are trained to learn a mapping function between several input features and the output value, which is denoted by the target to be predicted. Recent evidences reveal that DL methods have been promising tools for time series prediction, such as urban traffic flow \cite{LvITS2015,WuPartC2018}, cycle time of wafer lots \cite{WangLSTMTII2018}, wind speed \cite{ZhangTSE2015}, and financial market \cite{CavalcanteESA2016}, etc. However, DL-based prediction performance is highly dependent on the volume and diversity of training datasets. If the training datasets do not include the similar features existed in time series to be predicted, it becomes difficult to generate satisfactory results. The high computational complexity has also limited the deployability of DL-based prediction methods in practical applications. Therefore, to make prediction easier and more flexible, this paper will mainly focus on NN and its extensions for non-stationary time series prediction.

NN has the capacity of learning the potential complex relationships between different variables. Traditional NN methods, e.g., wavelet neural network (WNN) \cite{ChenNeuro2006}, fuzzy neural network (FNN) \cite{TangITS2017}, Elman neural network (ENN) \cite{ChandraNNLS2015}, back propagation neural network (BPNN) \cite{GaxiolaIS2014}, and generalized regression neural network (GRNN) \cite{SpechtTNN1991}, etc., have attracted considerable attention over the past decades. The development of maritime time series is often affected by several complex influences. It is still difficult to generate satisfactory prediction results since NN methods have their own disadvantages. For example, BPNN can fall into the local minimum problem during training process; GRNN often suffers from high computational time. To further improve prediction performance, empirical mode decomposition (EMD) and NN have been combined to contribute a two-step prediction framework \cite{RenTSE2015,WangNN2017}. For example, the combination versions, such as EMD+WNN \cite{SanthoshECM2018}, EMD+FNN \cite{WeiASC2016}, EMD+ENN \cite{WangASC2014}, EMD+BPNN \cite{WangRE2016,WeiChenPartC2012}, EMD+GRNN \cite{ZhouSTE2014} perform well in time series prediction. The reason behind this phenomenon is that EMD can decompose original non-stationary time series into a series of independent and nearly periodic components which admit the well-behaved Hilbert transforms. Each component, which is related to an Intrinsic Mode Function (IMF), can reveal hidden patterns and trends of time series. Especially, the IMF component with the lowest frequency generally denotes the trend or mean of original time series \cite{LangACCESS2018}. This decomposition strategy is thus able to effectively assist in developing prediction methods. In particular, EMD-based high- and low-frequency components are respectively predicted using traditional NN methods (e.g., WNN, FNN, ENN, BPNN and GRNN). The final results are obtained by aggregating the predicted high- and low-frequency components.
\subsection{Motivation and Contributions}
It is generally thought that original time series are commonly composed of several high- and low-frequency components which correspond to different characteristics of time series. The low-frequency components with stable change can be accurately predicted directly using WNN \cite{SanthoshECM2018}, FNN \cite{WeiASC2016}, ENN \cite{WangASC2014}, BPNN \cite{WangRE2016,WeiChenPartC2012}, and GRNN \cite{ZhouSTE2014} in the popular two-step prediction framework \cite{RenTSE2015,WangNN2017}. However, it becomes difficult to predict high-frequency components due to their properties of unobvious mathematical regularity. NN-based prediction results may lead to non-optimal results in practical applications. We find that the self-similarities\footnote{It means that each high-frequency component theoretically contains several small (overlapping) segments which have the similar geometrical shapes.} within high-frequency components can enable robust and accurate predictions. In our proposed three-step prediction framework, the high-frequency components are first divided into several continuous small (overlapping) segments. The geometric similarity between any two different segments can be effectively measured using the dynamic time warping (DTW) method \cite{MeiToC2016}. Compared with standardized euclidean distance (SED)-based similarity measures \cite{LiICONIP2018}, DTW is able to find the points with similar shapes to improve the similarity measures between different segments. The grouped segments with high DTW similarities are then selected to form the training datasets for traditional NN methods. It is able to enhance the predictions of high-frequency components. All the predicted components are combined as the final prediction of our proposed method.

In conclusion, our major contributions, given the current state-of-the-art research work, are mainly summarized by the following three aspects:
\begin{enumerate}
	\item The unified three-step prediction frameworks (called EMD+DTW+NN and EEMD+DTW+EE) are presented for non-stationary time series prediction in maritime industries. The satisfactory prediction performance benefits from the EMD-or EEMD-based property decomposition and DTW-based similarity grouping.
	\item Non-stationary time series are decomposed into high- and low-frequency components. The inherent self-similarities within high-frequency components are taken into account in promoting prediction accuracy. The predictions of low-frequency components are performed directly using traditional NN methods.
	\item Comprehensive experiments on port cargo throughput and vessel traffic flow have illustrated the superior performance of the proposed three-step prediction framework in terms of prediction accuracy and robustness.
\end{enumerate}

The main benefit of our proposed prediction methods is that it takes full advantage of the EMD- or EEMD-based property decomposition and DTW-based similarity grouping. Therefore, the prediction accuracy and robustness can be effectively enhanced among different applications. 
\subsection{Organization}
The remainder of this paper is organized as follows. The decomposition-based prediction frameworks are proposed in Section \ref{section2}. Section \ref{section3} is dedicated to several numerical experiments on both port cargo throughput and vessel traffic flow prediction. This paper is concluded by summarizing our main contributions in Section \ref{section4}.
\begin{figure*}[t]
	\centering
	\includegraphics[width=\linewidth]{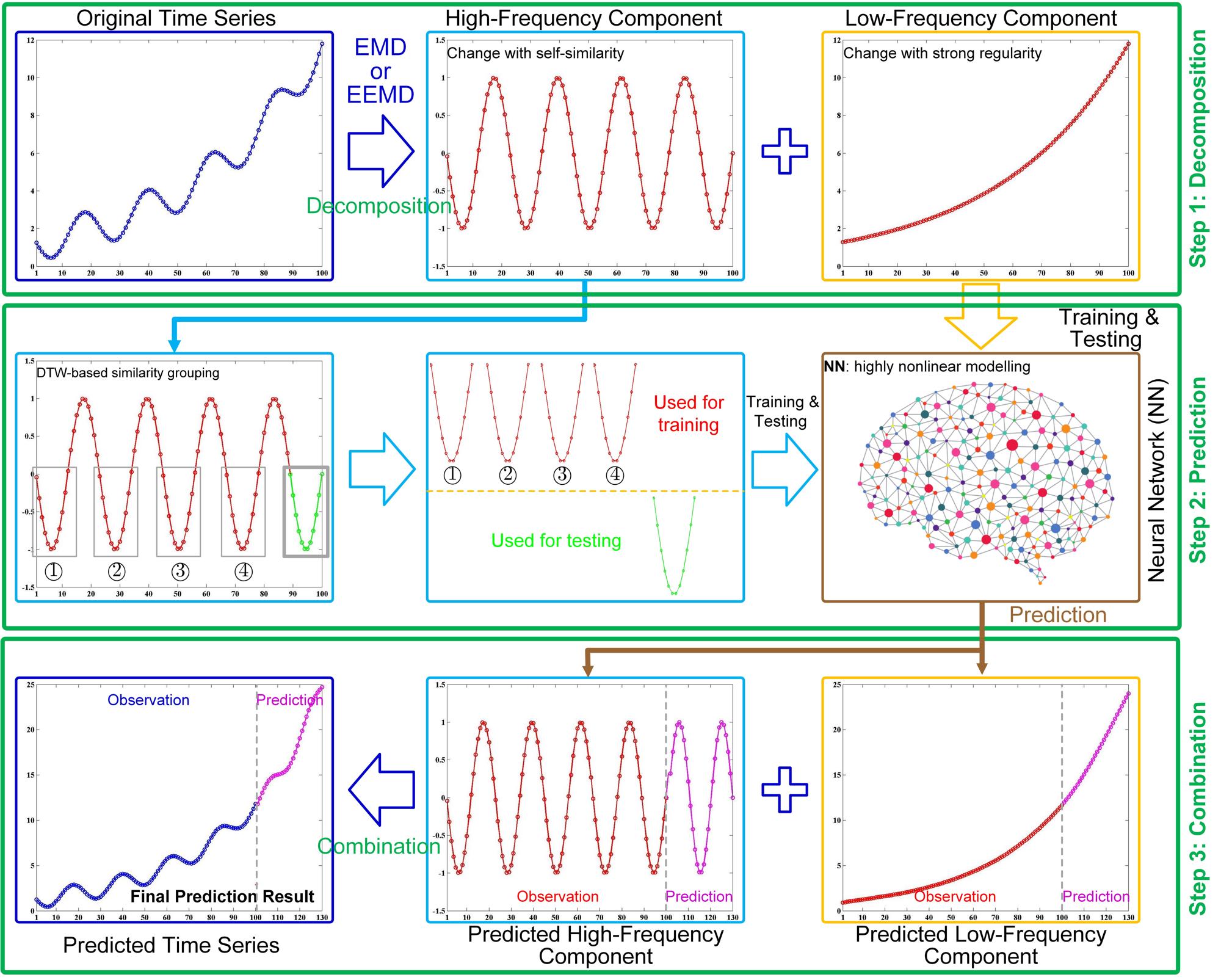}
	\caption{Flowchart of the proposed three-step prediction frameworks EMD+DTW+NN and EEMD+DTW+NN. For the sake of simplification, the original time series are decomposed into only two components. In practice, more than two components should be generated to guarantee high-accuracy prediction.}
	\label{Figure1}
\end{figure*}
\begin{figure*}[t]
	\centering
	\includegraphics[width=\linewidth]{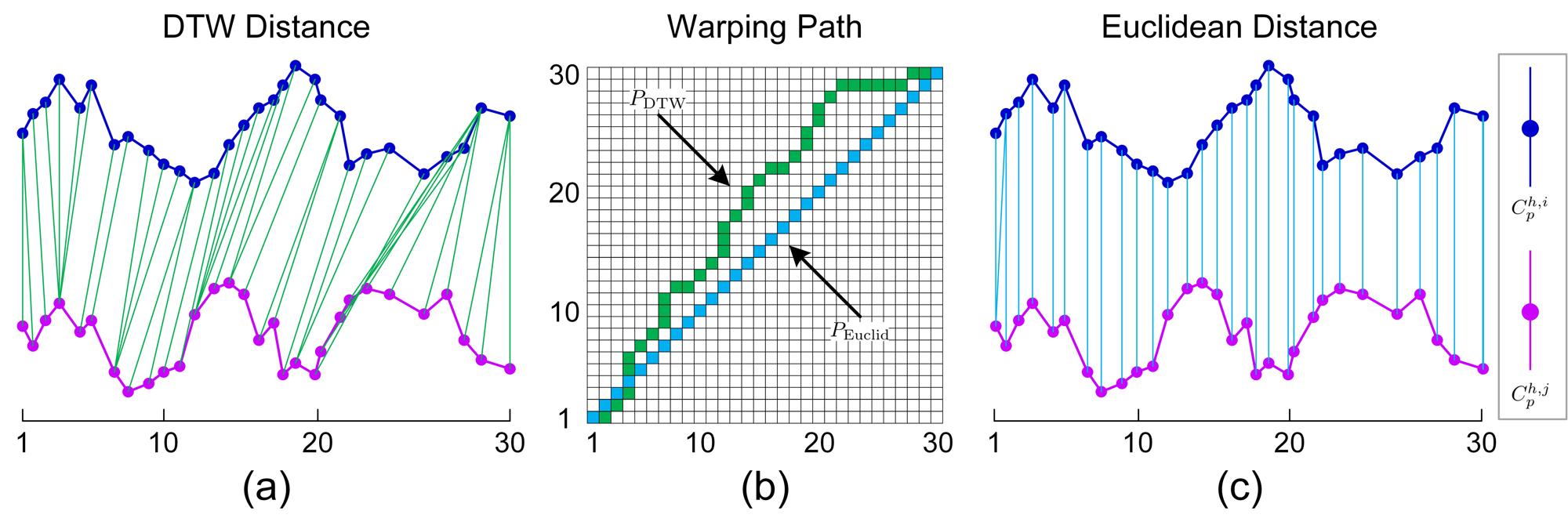}
	\caption{The visual illustration of differences between DTW and Euclidean distances. From left to right: (a) displays the DTW-based alignments between two time series for measuring distance (similarity), (b) shows the warping paths $P_{\mathrm{DTW}}$ and $P_{\mathrm{Euclid}}$, respectively, generated by calculating DTW and Euclidean distances, and (c) displays the matching between two time series using Euclidean distance. It can be found that DTW performs better in finding the points with similar geometric shapes to enhance the accuracy of distance (similarity) measure \cite{LiDTW2015}.}
	\label{FigureDTWDistance}
\end{figure*}
\section{Similarity Grouping-Guided Neural Network Modeling}
\label{section2}
This section will detailedly illustrate our proposed three-step prediction framework, shown in Fig. \ref{Figure1}. In the first step, EMD or EEMD can be adopted to decompose the original non-stationary time series into high- and low-frequency components. In the second step, the low-frequency components are effectively predicted directly using traditional NN methods. In contrast, the high-frequency components are first divided into several continuous small (overlapping) segments. Several segments with high DTW-based similarities are then grouped to form new training datasets for traditional NN methods to improve prediction accuracy. The predicted high- and low-frequency components are combined to form the final prediction results in the final step.
\subsection{EMD-Based Time Series Decomposition}
\label{sec:EMD}
In the literature \cite{ChenTIE2016}, EMD has been applied to decompose the non-stationary signals (i.e., time series) into a small number of different scaled data sequences. Each sequence can be designated as an Intrinsic Mode Function (IMF), which is independent and nearly periodic from a mathematical point of view. The implementation of DTW must meet the following two conditions
\begin{enumerate}
	\item The number of zero-crossing points and extrema points (local maxima and local minima) in the entire signal should be equal or differ by at most one.
	\item At any point in the signal, the mean values of lower and upper envelopes which are represented by the local minima and local maxima should be equal to zero.
\end{enumerate}

Essentially, EMD highly depends on the local characteristics of original signal, such as local minima, local maxima and zero-crossings. The decomposition of original non-stationary signals into IMFs was commonly performed using an iterative procedure called sifting algorithm. In particular, it calculated the IMFs at each scale from fine to coarse. For more details on EMD, we refer the interested reader to see \cite{FlandrinSPL2004} and references therein. Let $X(t)$ denote the original nonstationary time series, which can be decomposed as a sum of finite number of IMFs as follows
\begin{equation}\label{expressionIMF}
	{X(t) = \sum_{i=1}^{N} C_i(t) + R_{N}(t)},
\end{equation}
with $1 \leq t \leq T$. Here, $N$ is the total number of IMFs, $C_{i} (t)$ is the $i$th IMF, and $R_{N}(t)$ is the residual signal denoting the mean trend of original data $X(t)$. In practice, $R_{N}(t)$ can also be considered as an IMF and denoted as $C_{N+1}(t)$. Therefore, Eq. (\ref{expressionIMF}) can be rewritten as $X(t) = \sum_{i=1}^{N+1} C_i(t)$. In this decomposition, the first IMF is related to the fastest fluctuating part of $X(t)$; meanwhile the last IMF corresponds to the slowest fluctuating part.
\begin{figure*}[t]
	\centering
	\includegraphics[width=\linewidth]{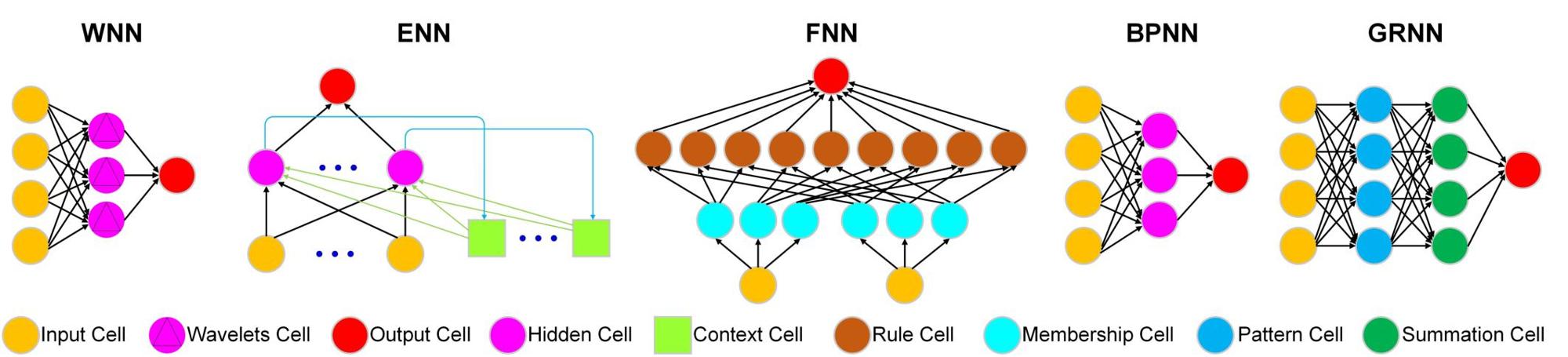}
	\caption{The architectural diagrams of $5$ different NN methods adopted in this paper. \textbf{WNN}: Wavelet Neural Network, \textbf{ENN}: Elman Neural Network, \textbf{FNN}: Fuzzy Neural Network, \textbf{BPNN}: Back Propagation Neural Network, and \textbf{GRNN}: Generalized Regression Neural Network.}
	\label{Fig:NNmethods}
\end{figure*}
%
%
%
%
\subsection{EEMD-Based Time Series Decomposition}
\label{sec:EEMD}
Traditional EMD, introduced in Section \ref{sec:EMD}, easily suffers from the mode mixing problem. This problem easily causes serious aliasing in the time-frequency distribution leading to degrading the decomposition accuracy. To promote the decomposition performance, Wu and Huang \cite{HuangEEMD2009} proposed to develop a new noise-assisted method named EEMD by performing the same iterative sifting process done in EMD. It repeatedly decomposes the time-series data into different IMFs via traditional EMD method. In particular, the IMF components in EEMD are defined as the ensemble of trails mean. Each trail consists of decomposition results of time-series data added by a uniformly distributed white noise of finite amplitude. Both theoretical and practical results have demonstrated that the added noise could help data analysis in the EMD method  \cite{WangJAG2012}. The ensemble means of the corresponding IMFs calculated from each trial are subsequently regarded as the IMFs of EEMD \cite{HuangEEMD2009}. The detailed procedures of EEMD are given as follows
	\begin{enumerate}
		\item Add a uniformly distributed white noise series to the original time-series data.
		\item Decompose the time-series data with added white noise into different IMFs via EMD shown in Section \ref{sec:EMD}.
		\item Iteratively repeat Steps (1) and (2) with different white noises and yield the corresponding IMF components. The ensemble number is essentially related to the number of repeated procedures.
		\item Calculate the mean of the ensemble IMFs as the final decomposition results.
	\end{enumerate}

For both EMD- and EEMD-based time series decomposition, IMFs can be considered as the combination of high- and low-frequency components in this work. The low-frequency components can be easily predicted directly using traditional NN methods in practice. In contrast, it is difficult to predict the high-frequency components due to their properties of weak mathematical regularity.
\subsection{DTW-Based Similarity Grouping}
\label{sec:DTW}
DTW is an effective and popular technique for measuring distance (or similarity) between two sequences of time series, which is based on the dynamic programming approach. It has been widely applied to voice identification, data clustering, feature extraction, etc. The similarity between two time series sequences is inversely proportional to the geometrical distance. The basic principle of DTW algorithm is to compare two time sequences and measure their similarities by computing a minimum distance between these two time series.

Let $Y = \left\{ y_{1}, y_{2}, \dots, y_{m} \right\}$ and $Z = \left\{ z_{1}, z_{2}, \dots, z_{n} \right\}$ denote two time sequences. Before the calculation of distance, we first create a $m \times n$ patch matrix where the $\left( i\mathrm{th}, j\mathrm{th} \right)$ element represents the distance $d \left( y_{i}, z_{j} \right)$ between the two points $y_{i}$ and $z_{j}$. To enhance computational robustness, $d \left( y_{i}, z_{j} \right)$ corresponds to the weighted Euclidean distance. Therefore, the DTW-based similarity is accordingly insensitive to the noise or outliers. As shown in Fig. \ref{Figure1}, it is easy to obtain that $m = n$ in our work. The best match between these two sequences $Y$ and $Z$ corresponds to the shortest distance path aligning one sequence to the other. Therefore, the optimal warping patch can be recursively calculated by
\begin{equation}\label{Eq:DTW}
	\mathrm{DTW} \left( Y, Z \right) = \gamma \left( i, j \right),
\end{equation}
where the minimum cumulative distance $\gamma \left( i, j \right)$ is given by
\begin{align}\label{Eq:expression2}
	\gamma \left( i, j \right) &= d \left( y_{i}, z_{j} \right) \\
	&+ \min \left\{ \gamma \left( i-1, j-1 \right), \gamma \left( i-1, j \right), \gamma \left( i, j-1 \right) \right\}. \nonumber
\end{align}

For the sake of better understanding, the visual illustration of differences between DTW and Euclidean distances is shown in Fig. \ref{FigureDTWDistance}. Compared with the widely-used Euclidean distance, DTW performs better in extracting the points with similar geometric shapes leading to enhancing the accuracy of distance (similarity) measure \cite{LiDTW2015}. It is worth noting that high-frequency components (i.e., IMFs) are difficult to predict due to their properties of high levels of volatility. In this work, high-frequency IMFs will be divided into a series of overlapping segments. The segments with high similarity, measured using DTW distance, are merged into the same group. The clustered groups are exploited to form more proper training datasets for traditional NN methods to enhance prediction performance.
\begin{figure*}[t]
	\centering
	\includegraphics[width=\linewidth]{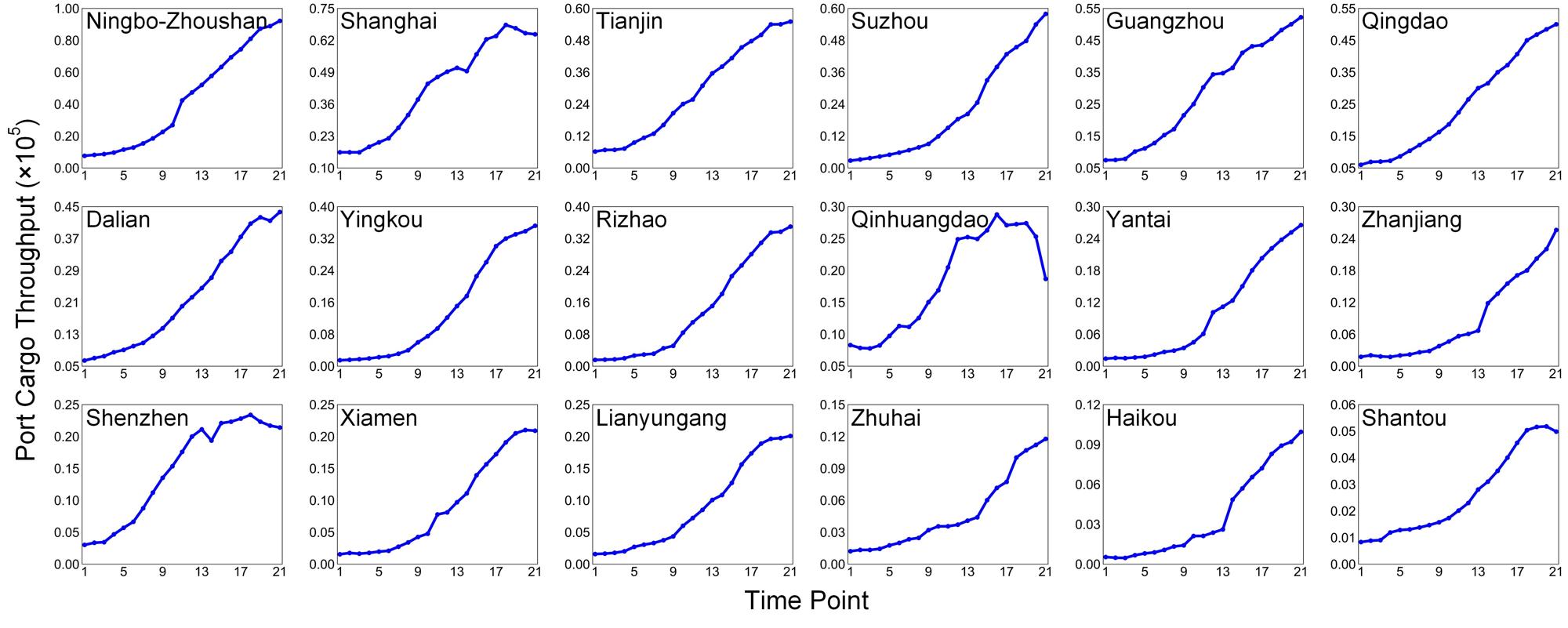}
	\caption{The growing trends of PCT time series data at 18 major ports in China from 1996 to 2016 (i.e., \textit{Time Point from $1$ to $21$ in the horizontal axis}). It can be found that different ports have different properties of growing trends.}
	\label{Figureoridata}
\end{figure*}
\subsection{Three-Step Prediction Framework}
The proposed three-step framework for non-stationary time series prediction is illustrated in Fig. \ref{Figure1}. The basic theories of EMD, EEMD and DTW have been introduced in Sections \ref{sec:EMD}-\ref{sec:DTW}. The detailed calculation steps for our EMD+DTW+NN and EEMD+DTW+NN are described as follows
\subsubsection{\textbf{Step 1--Decomposition}} EMD or EEMD method first decomposes the original non-stationary time series $X$ into high-frequency IMFs $C^{h}_{1 \leq p \leq P}$ and low-frequency IMFs $C^{l}_{1 \leq q \leq Q}$ with $P + Q = N + 1$. Although $C^{h}_{1 \leq p \leq P}$ can extract the main structures and characteristics in $X$, it is still difficult to predict $C^{h}_{1 \leq p \leq P}$ due to its weak mathematical regularity property. In contrast, it becomes easier to predict $C^{l}_{1 \leq q \leq Q}$ since it indicates the long-period characteristics. Note that the last IMF $C^{l}_{Q}$ generally represents the essential trend of $X$ and its high-accuracy prediction could be guaranteed. For the sake of simplicity, we tend to respectively adopt $C^{h}_{p}$ and $C^{l}_{q}$ instead of $C^{h}_{1 \leq p \leq P}$ and $C^{l}_{1 \leq q \leq Q}$ throughout the rest of this paper.
\subsubsection{\textbf{Step 2--Prediction}} The low-frequency IMFs $C^{l}_{q}$ will be directly predicted using traditional NN methods, e.g., WNN, FNN, ENN, BPNN and GRNN. The mathematical foundations behind these five NN methods have been vastly discussed in current research work. For the sake of maximum conciseness and better understanding, only the corresponding architectural diagrams are introduced in this paper, visually illustrated in Fig. \ref{Fig:NNmethods}. It is more difficult to generate the accurate predictions of high-frequency IMFs $C^{h}_{p}$ because of their properties of high levels of volatility. In order to overcome this shortcoming, we propose to take advantage of the self-similarities within $C^{h}_{p}$. In particular, each $C^{h}_{p}$ is first divided into a series of overlapping segments, i.e., $\mathcal{C}^{h}_{p} = \{ \{ C^{h}_{p} ( 1 ), C^{h}_{p} ( 2 ), \cdots, C^{h}_{p} ( L ) \}, \cdots, \{ C^{h}_{p} ( i ), C^{h}_{p} ( i+1 ), \cdots, C^{h}_{p} ( L+i-1 ) \}, \cdots, \{ C^{h}_{p} ( T - L + 1 ), C^{h}_{p} ( T - L + 2 ), \cdots, C^{h}_{p} ( T ) \} \}$, with $L$ denoting the length of each segment. How to construct the appropriate training dataset adopted for NN methods significantly effects the final prediction performance. Taking the prediction of $C^{h}_{p} ( T + 1 )$ as an example, we consider $C^{h,T-L+1}_{p} = \{ C^{h}_{p} ( T - L + 1 ), C^{h}_{p} ( T - L + 2 ), \cdots, C^{h}_{p} ( T ) \}$ as an input reference segment. The similarities between $C^{h,T-L+1}_{p}$ and other segments in $\mathcal{C}^{h}_{p}$ are then calculated using the DTW method, shown in Fig. \ref{FigureDTWDistance}. DTW has the capacity of generating robust estimation of similarity between different segments. 

In this work, we propose to automatically group the segments with higher similarities from $\mathcal{C}^{h}_{p}$ to form more appropriate training dataset for NN-based prediction of high-frequency components $C^{h}_{p} ( T + 1 )$. Analogous to this strategy, the accurate prediction of $C^{h}_{p} \left( T + \cdot \right)$ could be performed accordingly. It is worth noting that some segments in $\mathcal{C}^{h}_{p}$ are inappropriate for promoting prediction if there are low degrees of similarity between these segments and the reference one. Within our framework, only the grouped segments with high degrees of similarity are extracted to enhance prediction accuracy and robustness.
\begin{figure*}[t]
	\centering
	\includegraphics[width=\linewidth]{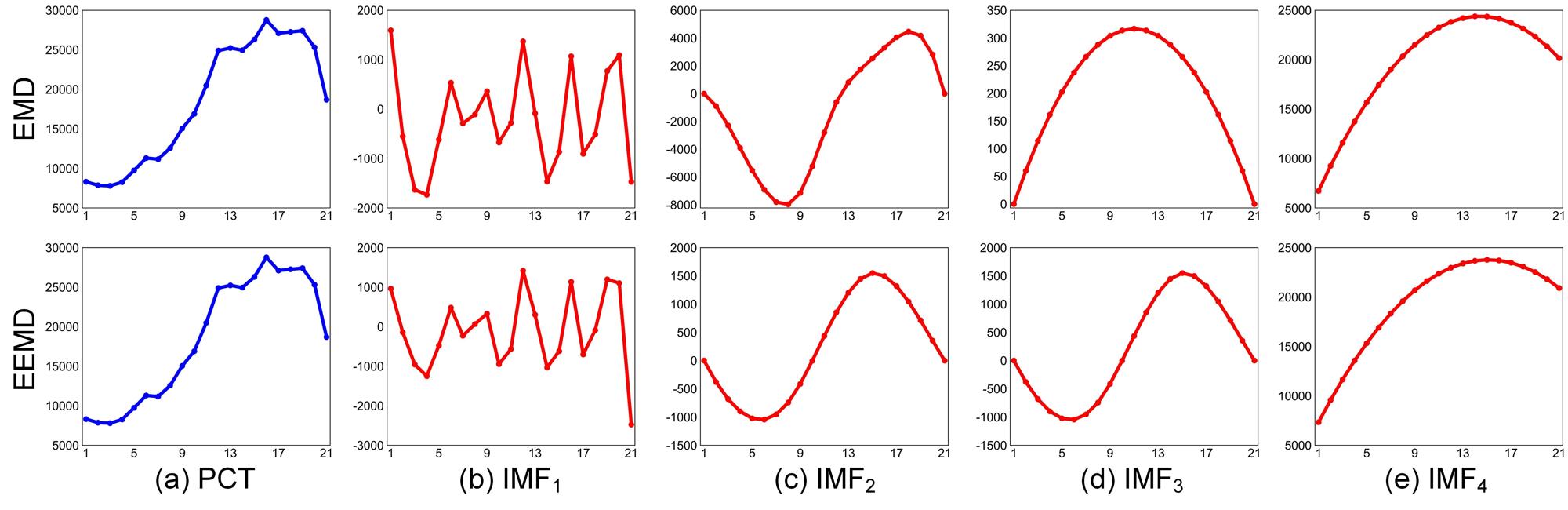}
	\caption{EMD- and EEMD-based decompositions of PCT time series for an example of Qinghuangdao Port. From left to right: (a) original PCT time series, (b) IMF$_{1}$, (c) IMF$_{2}$, (d) IMF$_{3}$ and (e) IMF$_{4}$. In particular, IMF$_{1}$ corresponds to the high-frequency component difficult to predict in this work. IMF$_{2}$, IMF$_{3}$ and IMF$_{4}$ are the low-frequency components easy to deal with in practical applications.}
	\label{FigureEMD}
\end{figure*}
\subsubsection{\textbf{Step 3--Combination}} High-frequency IMFs $C^{h}_{1 \leq p \leq P}$ are predicted using our proposed EMD+DTW+NN and EEMD+DTW+NN to enhance prediction accuracy; whereas low-frequency IMFs $C^{l}_{1 \leq q \leq Q}$ are predicted directly using traditional NN methods to shorten computational time. It is worth mentioning that the prediction accuracy of low-frequency IMFs could be further improved via EMD+DTW+NN and EEMD+DTW+NN but at the expense of much higher computational load. In this work, the final prediction result is obtained by integrating the predicted high- and low-frequency IMFs, i.e.,
\begin{equation}\label{Eq:predictionresults}
	{X \left( T + \hat{t} \right) = \sum_{p=1}^{P} C^{h}_{p} \left( T + \hat{t} \right) + \sum_{q=1}^{Q} C^{l}_{q} \left( T + \hat{t} \right)},
\end{equation}
for $1 \leq \hat{t} \leq L_{T}$ with $L_{T}$ being the length of time-series data points needed to be predicted in numerical experiments. Through the proposed two three-step prediction frameworks, it is able to generate more accurate and stable prediction performance.
\section{Experimental Results and Discussion}
\label{section3}
To evaluate the prediction accuracy and robustness, our proposed three-step prediction frameworks will be compared with several popular methods on realistic datasets in maritime industries, e.g., port cargo throughput and vessel traffic flow. The Matlab source codes of our three-step prediction frameworks can be download at \textcolor[rgb]{0,0,1}{\url{http://mipc.whut.edu.cn/}}.
\subsection{Non-Stationary Time Series in Maritime Industry}
\subsubsection{Port Cargo Throughput (PCT)}
In this work, we select the realistic PCT time series from $18$ major ports in China from $1996$ to $2016$. The growing trends of PCT time series are visually illustrated in Fig. \ref{Figureoridata}. It shows that different ports have different properties of growing trends due to several influence factors, such as geographic location, type of goods, port scale, service level, and management level, etc. It means that the proposed prediction method should have the universality and robustness properties to enhance prediction results.

\begin{figure*}[t]
	\centering
	\includegraphics[width=\linewidth]{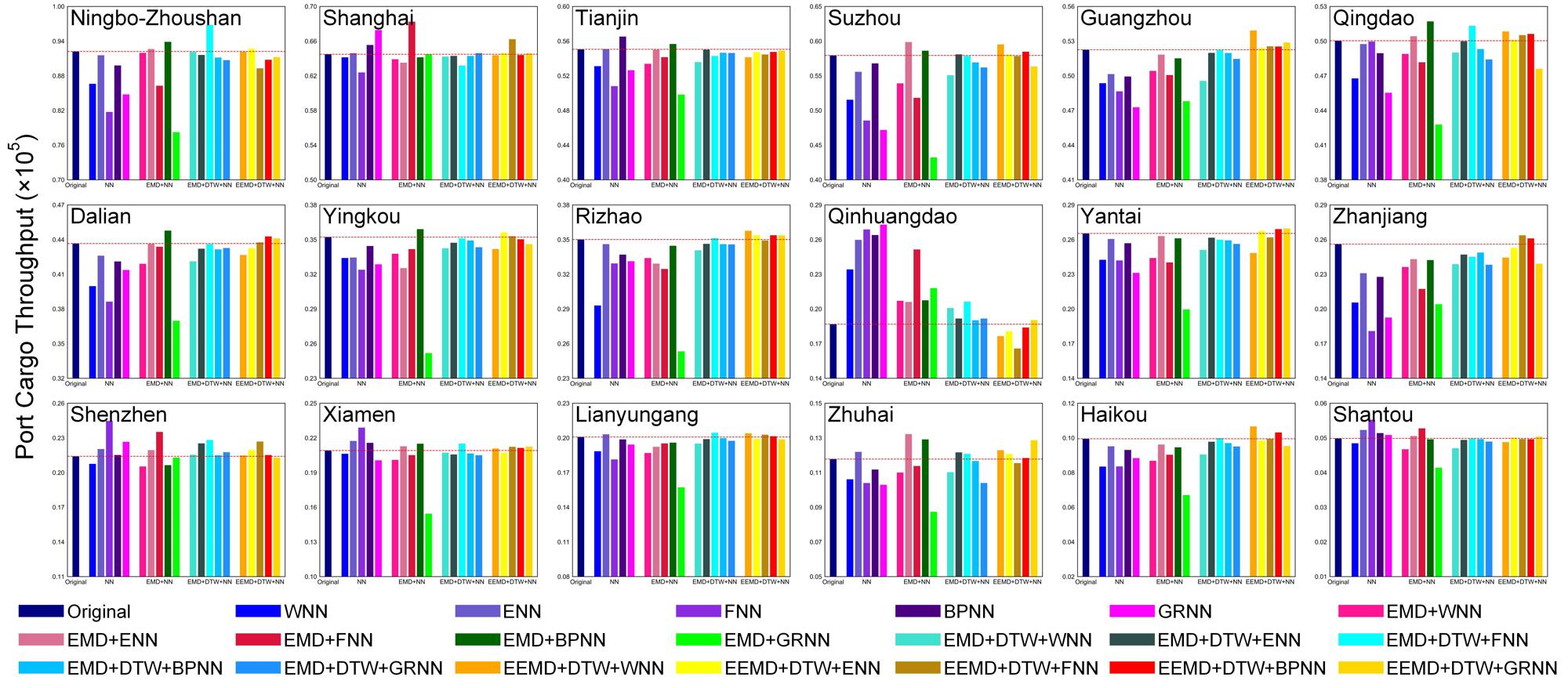}
	\caption{The PCT prediction results for $18$ major ports in China in $2016$. In each subfigure, the adopted prediction methods, from left to right, are NN (i.e., WNN, ENN, FNN, BPNN, GRNN), EMD+NN (i.e., EMD+WNN, EMD+ENN, EMD+FNN, EMD+BPNN, EMD+GRNN), EMD+DTW+NN (i.e., EMD+DTW+WNN, EMD+DTW+ENN, EMD+DTW+FNN, EMD+DTW+BPNN, EMD+DTW+GRNN), and EEMD+DTW+NN (i.e., EEMD+DTW+WNN, EEMD+DTW+ENN, EEMD+DTW+FNN, EEMD+DTW+BPNN, EEMD+DTW+GRNN), respectively. The dashed lines in subfigures indicate the actual PCT data in $2016$.}
	\label{FigurePred2016}
\end{figure*}
\subsubsection{Vessel Traffic Flow (VTF)}
We propose to adopt the original VTF time series at the Wuhan Yangtze Great Bridge to evaluate the medium-term prediction performance. The original VTF data can be generated using the dynamic automatic identification system (AIS) information\footnote{The AIS, an automatic tracking system used on ships, is able to provide real-time vessel position (latitude and longitude), course and speed, etc.}. In particular, the VTF data of $16$ days are extracted at $3$-hour intervals to assist in medium-term prediction. The original VTF data can be found in Fig. \ref{Fig:VTFDecomposition} in Section \ref{sec:VTFprediction}. To the best of our knowledge, previous work mainly focuses on long-term VTF prediction. In contrast, our VTF time series have the property of higher temporal resolution compared with previous research. The medium-short prediction results are beneficial to delicacy management and traffic safety enhancement in maritime industries.
\subsection{Comparison with Other Prediction Methods}
Our proposed three-step frameworks EMD+DTW+NN and EEMD+DTW+NN will be mainly compared with traditional prediction methods (i.e., NN and EMD+NN) as follows
\begin{itemize}
	\item \textbf{NN}: Without loss of generality, widely-used NN methods, such as WNN, FNN, ENN, BPNN and GRNN, are introduced to directly predict non-stationary time series in maritime industries. Both EMD+NN and EMD+DTW+NN frameworks are developed based on these $5$ popular NN methods to further enhance prediction performance.
	\item \textbf{EMD+NN}: EMD+NN is essentially a two-step prediction framework which combines EMD with traditional NN methods. As done in \cite{WangNN2017,RenTSE2015,RenNNLS2016}, the original non-stationary time series are first decomposed into several high- and low-frequency components using the EMD method. The NN methods are then exploited to directly predict both high- and low-frequency components. The final prediction results are obtained by combining the predicted high- and low-frequency components. Theoretically, EMD+NN has the capacity of guaranteeing more robust performance compared with single NN methods.
	\item \textbf{EMD+DTW+NN}: This three-step prediction framework can be considered as an extension of EMD+NN through taking full advantage of self-similarity features within high-frequency components. Under this framework, the prediction of low-frequency components will be directly performed using the popular NN methods. In contrast, the high-frequency components are divided into several small segments which are grouped via DTW-based similarity measure. The grouped segments are selected to form more proper training datasets for NN to enhance prediction accuracy. Both predicted high- and low-frequency components are also combined to generate the final prediction results.
	\item \textbf{EEMD+DTW+NN}: This prediction framework is essentially an extension of EMD+DTW+NN by replacing EMD with EEMD during time series decomposition. EEMD is a noise-assisted data analysis method \cite{HuangEEMD2009}, which generates more robust and stable decomposition results compared with simple EMD. Theoretically, EEMD-based three-step calculation framework has the capacity of yielding superior prediction results.
\end{itemize}

To quantitatively analyze the prediction results, relative error (RE) is selected as a measure of accuracy. The corresponding mathematical formulation is defined as follows
\begin{equation}
	\mathrm{RE} = \frac{|y - \widehat{y}|}{y},
\end{equation}
with $y$ and $\widehat{y}$, respectively, being the actual and predicted time series. To further evaluate the prediction robustness, each prediction method will run $10$ times to generate the mean and standard deviation versions of RE in our numerical experiments.
\begin{figure*}[t]
	\centering
	\includegraphics[width=\linewidth]{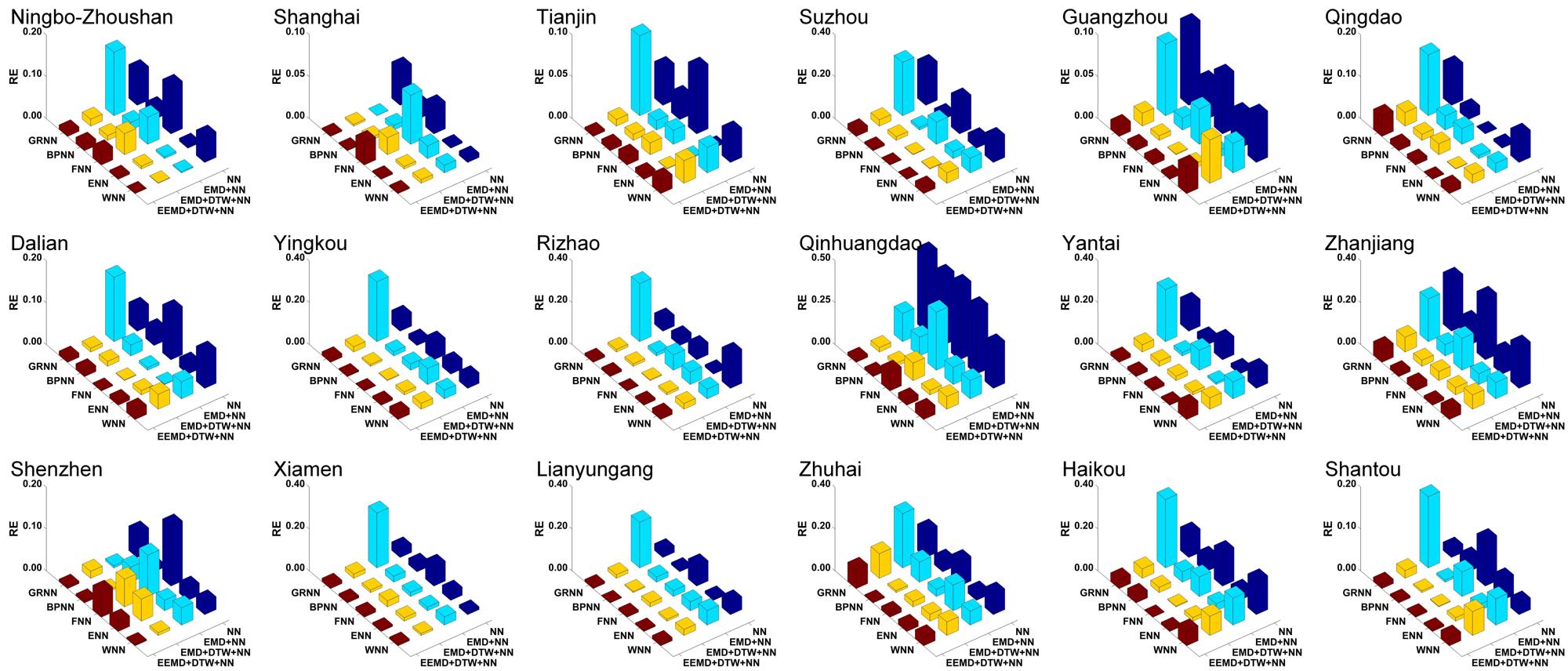}
	\caption{Visual illustration of mean RE for different prediction frameworks at 18 major ports in China. To reduce randomness, each prediction method ran $10$ times to obtain the average results. Results demonstrate that our proposed frameworks yield the lowest RE under all conditions.}
	\label{FigureRE}
\end{figure*}
\begin{figure*}[!]
	\centering
	\includegraphics[width=\linewidth]{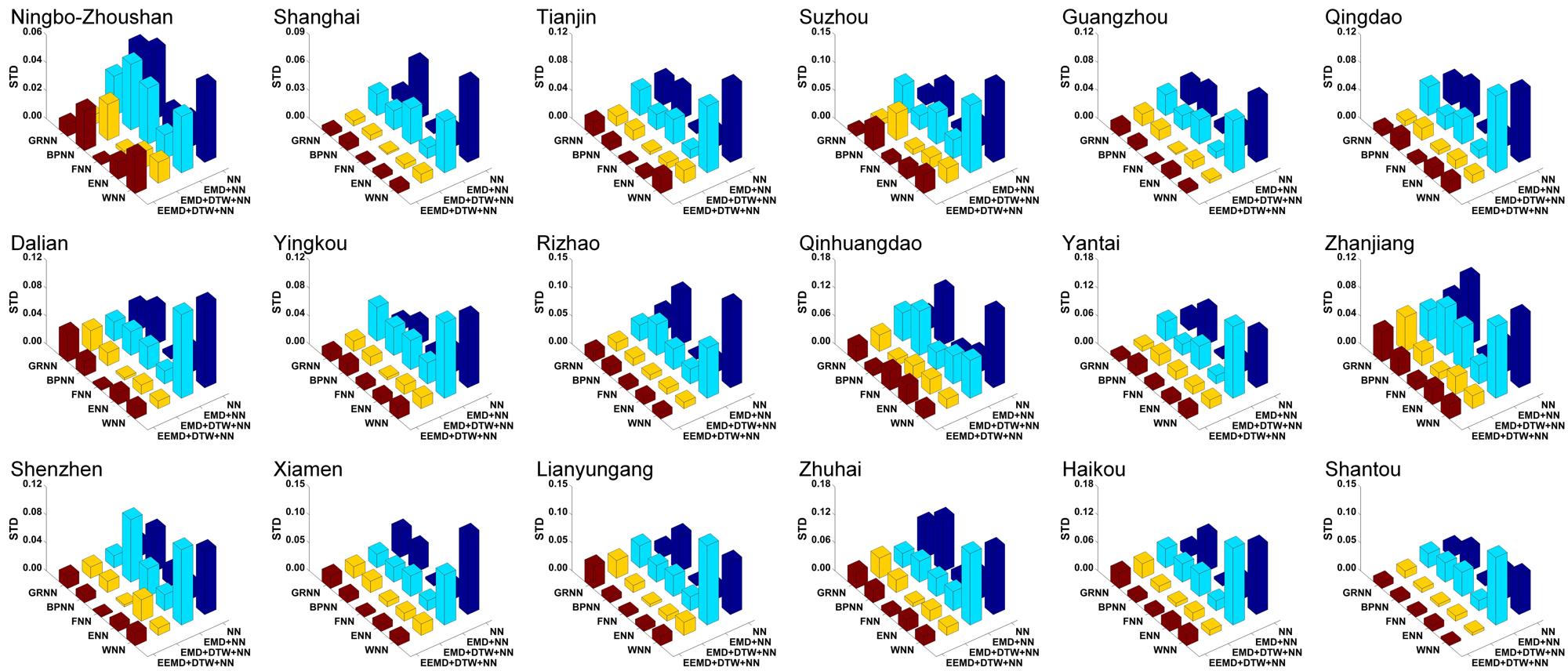}
	\caption{The display of standard deviation of RE for different prediction methods at 18 major ports in China. This standard deviation is adopted to evaluate the prediction robustness. To reduce randomness, each prediction method ran $10$ times to obtain the standard deviation of RE. Statistical results illustrate that our proposed frameworks generate the most robust prediction under consideration in most of the cases.}
	\label{FigureVarianceRE}
\end{figure*}
\begin{figure*}[!]
	\centering
	\includegraphics[width=\linewidth]{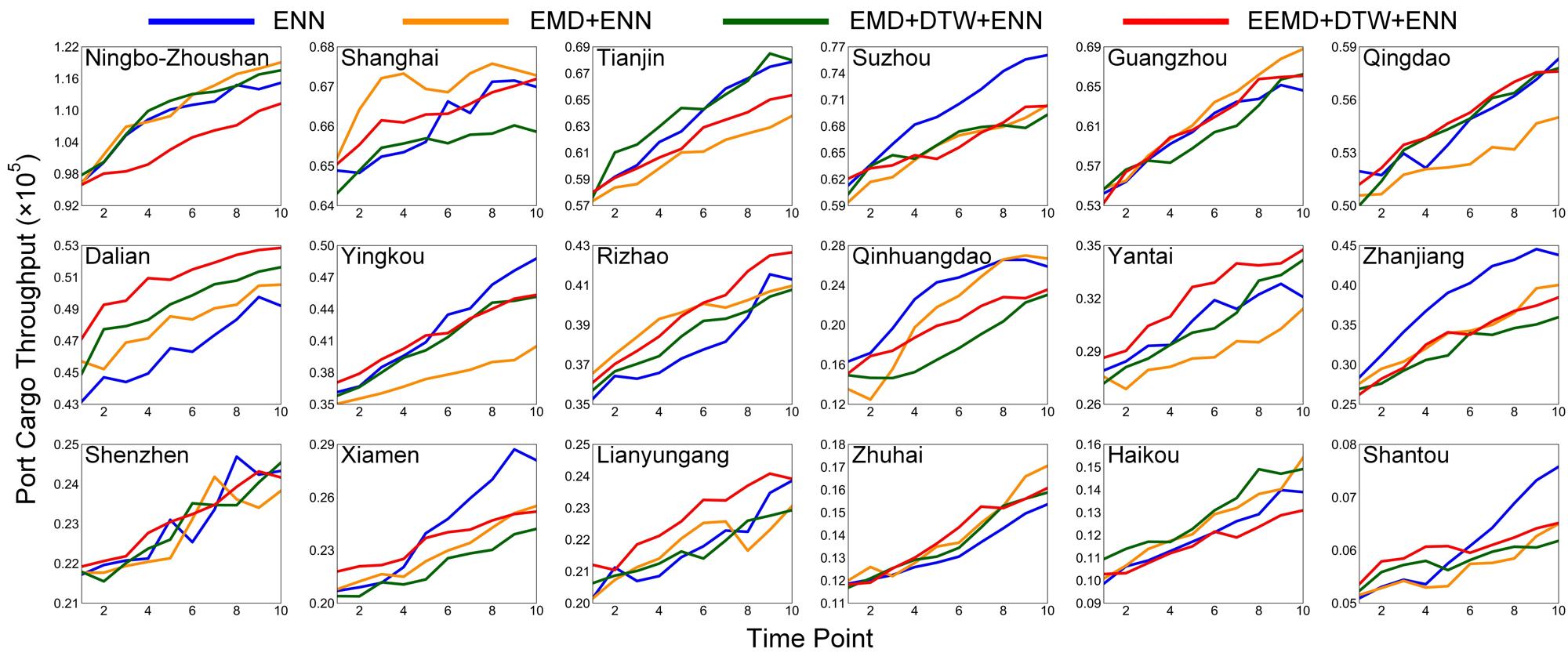}
	\caption{The PCT prediction results for $18$ major ports in China from $2017$ to $2026$ (i.e., \textit{Time Point from $1$ to $10$ in the horizontal axis}). It can be found that the proposed EMD+DTW+ENN and EMD+DTW+ENN frameworks are able to more robustly implement PCT prediction in practical applications.}
	\label{FigurePreResult}
\end{figure*}
\subsection{Experiments on Port Cargo Throughput Prediction}
\subsubsection{Quantitative Performance Evaluation}
To implement the experiments, the PCT time series in Fig. \ref{Figureoridata}, extracted at $18$ major ports from $1996$ to $2015$ in China, are used to train our three-step prediction frameworks. The prediction accuracy is evaluated using the predicted data from all these ports in $2016$. It can be found that the growing trend of PCT time series in Qinghuangdao Port appears more complicated. From a mathematical point of view, owing to the obvious regular changes, it is easier to accurately predict the PCT time series in other ports by directly using traditional techniques, e.g., Kalman filtering \cite{ZuluagaAE2015}, (non-)parametric regression \cite{DonnellyAE2015}, Wavelet filtering \cite{JooESA2015}, NN methods \cite{SanthoshECM2018, WeiASC2016, WangASC2014, WangRE2016, WeiChenPartC2012, ZhouSTE2014}, etc. Therefore, taking Qinghuangdao Port as an example, we propose to decompose the PCT time series into $4$ different components, shown in Fig. \ref{FigureEMD}. The first one is usually related to high-frequency IMF difficult to predict. The other $3$ components correspond to low-frequency IMFs. This similar decomposition manner will also be performed for other ports in this work.

In our numerical experiments, we empirically select $P = 1$ and $Q = 3$ in Eq. (\ref{Eq:predictionresults}) since extensive experimental results have demonstrated that this selection is able to generate satisfactory performance. The quantitative results from our proposed frameworks EMD+DTW+NN, EEMD+DTW+NN and other competing techniques (i.e., NN and EMD+NN) are detailedly depicted in Fig. \ref{FigurePred2016}. It can be observed that the direct utilization of NN generates the lowest prediction accuracy. In contrast, the combination of EMD and NN is able to improve prediction performance. For Qinghuangdao Port with more complicated trend of PCT data, our proposed methods EMD+DTW+NN and EEMD+DTW+NN generate superior prediction results compared with NN and EMD+NN. This is because of the fact that EMD- and EEMD-based PCT decomposition yields relatively stationary components which can be readily modeled using NN methods. It is well known that DTW can robustly measure the similarities between different small segments within decomposed components. The segments with high similarities are selected to group the training dataset for NN methods to enhance prediction performance. Therefore, our proposed two three-step frameworks can significantly outperform other competing methods. The corresponding prediction results, shown in Fig. \ref{FigurePred2016}, are closer to the dashed lines which indicate the actual PCT data. Since EEMD is essentially an improved version of EMD, EMD+DTW+NN produce predicted results which are more similar to the actual data.

The superior performance of our proposed frameworks is further confirmed by the statistical results for prediction accuracy and robustness shown in Figs. \ref{FigureRE} and \ref{FigureVarianceRE}. The mean RE results in Fig. \ref{FigureRE} illustrate that EMD+DTW+NN and EEMD+DTW+NN generate the most accurate predictions under all conditions. In contrast, EMD+NN has constrained the further improvement in prediction performance. Since the PCT time series have the essential property of weak mathematical regularity, single NN methods generate the lowest prediction accuracy under consideration in most of the cases. Fig. \ref{FigureVarianceRE} visually illustrates the standard deviation (std) of RE\footnote{The standard deviation of RE is essentially related to the robustness of prediction performance. In particular, low standard deviation indicates the robust prediction in practice.} adopted to evaluate the prediction robustness. It can be found that both EMD+DTW+NN and EEMD+DTW+NN can generate the most robust prediction. This good performance mainly benefits from the EMD- or EEMD-based property decomposition and DTW-based similarity grouping. 
\setlength{\tabcolsep}{2.5pt}
\begin{table*}
	\centering
	\caption{Prediction results (mean$\pm$std) of VTF time series from time points $121$ to $128$ for different methods (i.e., NN, EMD+NN, EMD+DTW+NN and EEMD+DTW+NN) at the Wuhan Yangtze Great Bridge. RE is used to quantitatively evaluate the prediction performance.}
	\begin{tabular}{|c|c|c|c|c|c|c|c|c|c|c|}
		\hline
		\multicolumn{2}{|l|}{Time Point}         & $121(03:00)$   & $122(06:00)$   & $123(09:00)$   & $124(12:00)$   & $125(15:00)$   & $126(18:00)$   & $127(21:00)$   & $128(24:00)$ & \multirow{2}{*}{RE} \\ \cline{1-10}
		\multicolumn{2}{|l|}{Actual Data}        & $34.00$ & $37.00$ & $36.00$ & $41.00$ & $48.00$ & $39.00$ & $38.00$ & $34.00$ &  \\ \hline \hline
		\multirow{5}{*}{\tabincell{c}{NN}}
		& WNN  & $26.47\pm5.91$ & $27.47\pm6.34$ & $32.51\pm4.37$ & $46.62\pm5.43$ & $44.58\pm7.41$ & $47.49\pm6.62$ & $42.83\pm7.11$ & $30.71\pm3.64$ & $0.15\pm0.07$ \\
		& ENN  & $28.45\pm3.85$ & $29.32\pm4.15$ & $33.76\pm4.26$ & $44.58\pm3.72$ & $44.34\pm5.89$ & $45.57\pm5.86$ & $43.38\pm4.17$ & $30.26\pm3.45$ & $0.13\pm0.05$ \\
		& FNN  & $28.56\pm4.28$ & $31.27\pm4.55$ & $30.28\pm3.97$ & $46.36\pm5.15$ & $42.34\pm6.87$ & $46.14\pm3.49$ & $44.42\pm4.59$ & $30.56\pm3.34$ & $0.15\pm0.03$ \\
		& BPNN & $27.35\pm3.49$ & $28.26\pm3.73$ & $32.21\pm4.54$ & $44.34\pm5.99$ & $44.67\pm3.54$ & $45.23\pm4.05$ & $42.02\pm5.11$ & $31.71\pm3.47$ & $0.13\pm0.06$ \\
		& GRNN & $29.23\pm5.17$ & $30.54\pm4.46$ & $31.42\pm5.62$ & $45.21\pm4.11$ & $43.11\pm4.83$ & $44.04\pm3.19$ & $44.16\pm5.87$ & $28.92\pm4.41$ & $0.14\pm0.03$ \\ \hline \hline
		\multirow{5}{*}{\tabincell{c}{EMD \\ + \\NN}}
		& WNN  & $29.32\pm4.27$ & $31.24\pm4.51$ & $34.16\pm4.53$ & $38.27\pm3.28$ & $45.62\pm4.23$ & $45.43\pm4.74$ & $42.86\pm6.78$ & $31.92\pm2.89$ & $0.10\pm0.05$ \\
		& ENN  & $30.24\pm3.44$ & $32.16\pm3.86$ & $34.15\pm2.24$ & $43.64\pm3.28$ & $45.75\pm3.37$ & $43.24\pm4.19$ & $41.34\pm3.75$ & $32.21\pm2.16$ & $0.08\pm0.03$ \\
		& FNN  & $31.95\pm4.22$ & $34.47\pm3.59$ & $30.71\pm3.81$ & $44.49\pm4.68$ & $44.37\pm4.34$ & $42.61\pm2.54$ & $42.65\pm3.17$ & $31.21\pm1.53$ & $0.09\pm0.03$ \\
		& BPNN & $30.56\pm3.21$ & $33.96\pm3.68$ & $34.88\pm2.39$ & $44.97\pm4.23$ & $46.13\pm3.58$ & $41.41\pm2.13$ & $41.44\pm4.28$ & $32.23\pm2.76$ & $0.07\pm0.03$ \\
		& GRNN & $29.33\pm5.39$ & $34.77\pm3.87$ & $32.65\pm4.61$ & $43.45\pm3.37$ & $45.94\pm4.65$ & $42.98\pm2.62$ & $40.49\pm5.34$ & $36.44\pm3.17$ & $0.08\pm0.03$ \\ \hline \hline
		\multirow{5}{*}{\tabincell{c}{EMD \\ + \\DTW\\ + \\NN}}
		& WNN  & $32.82\pm2.13$ & $35.83\pm2.74$ & $\textcolor[rgb]{0,1,0}{36.81\pm2.13}$ & $42.92\pm1.95$ & $46.37\pm2.08$ & $\textcolor[rgb]{0,1,0}{38.67\pm3.51}$ & $39.23\pm4.22$ & $35.03\pm1.27$ & $\textcolor[rgb]{0,0,1}{0.03\pm0.04}$ \\
		& ENN  & $\textcolor[rgb]{0,1,0}{33.21\pm1.94}$ & $\textcolor[rgb]{0,0,1}{36.12\pm1.51}$ & $37.83\pm2.95$ & $42.75\pm1.47$ & $\textcolor[rgb]{0,1,0}{46.95\pm2.93}$ & $41.34\pm3.79$ & $40.59\pm1.49$ & $34.87\pm1.14$ & $0.04\pm0.02$ \\
		& FNN  & $31.21\pm2.24$ & $38.57\pm1.44$ & $34.23\pm2.89$ & $43.99\pm1.91$ & $50.81\pm2.62$ & $40.43\pm1.47$ & $\textcolor[rgb]{0,1,0}{37.54\pm2.02}$ & $35.26\pm1.36$ & $0.05\pm0.02$ \\
		& BPNN & $35.77\pm1.67$ & $39.08\pm2.35$ & $\textcolor[rgb]{1,0,0}{35.46\pm1.12}$ & $42.88\pm1.47$ & $\textcolor[rgb]{0,0,1}{49.08\pm2.04}$ & $40.19\pm2.55$ & $39.58\pm1.96$ & $35.03\pm0.98$ & $0.04\pm0.01$ \\
		& GRNN & $32.51\pm1,42$ & $38.58\pm2.45$ & $34.46\pm1.43$ & $\textcolor[rgb]{0,0,1}{42.08\pm1.97}$ & $50.48\pm2.67$ & $37.19\pm1.73$ & $38.98\pm3.11$ & $32.32\pm2.49$ & $0.04\pm0.01$ \\ \hline
		\multirow{5}{*}{\tabincell{c}{EEMD \\ + \\DTW\\ + \\NN}}
		& WNN  & $\textcolor[rgb]{1,0,0}{33.25\pm1.74}$ & $\textcolor[rgb]{0,1,0}{36.43\pm2.16}$ & $35.07\pm2.15$ & $39.22\pm2.59$ & $46.12\pm2.67$ & $\textcolor[rgb]{1,0,0}{38.70\pm3.80}$ & $37.11\pm3.98$ & $\textcolor[rgb]{0,0,1}{33.28\pm1.89}$ & $\textcolor[rgb]{1,0,0}{0.02\pm0.01}$ \\
		& ENN  & $\textcolor[rgb]{0,0,1}{33.15\pm1.54}$ & $\textcolor[rgb]{1,0,0}{36.44\pm1.45}$ & $35.02\pm2.41$ & $\textcolor[rgb]{0,1,0}{40.02\pm2.16}$ & $46.55\pm2.32$ & $37.81\pm3.24$ & $36.09\pm0.83$ & $\textcolor[rgb]{1,0,0}{33.62\pm1.60}$ & $\textcolor[rgb]{0,1,0}{0.03\pm0.01}$ \\
		& FNN  & $32.33\pm1.46$ & $35.89\pm1.96$ & $34.94\pm2.71$ & $38.19\pm2.10$ & $44.89\pm2.14$ & $38.08\pm1.68$ & $\textcolor[rgb]{1,0,0}{38.23\pm2.44}$ & $\textcolor[rgb]{0,1,0}{33.38\pm1.80}$ & $0.04\pm0.02$ \\
		& BPNN & $32.62\pm1.23$ & $35.54\pm3.21$ & $\textcolor[rgb]{0,0,1}{35.08\pm1.58}$ & $38.64\pm1.45$ & $\textcolor[rgb]{1,0,0}{48.96\pm2.20}$ & $\textcolor[rgb]{0,0,1}{39.78\pm2.02}$ & $36.50\pm1.88$ & $33.11\pm0.91$ & $\textcolor[rgb]{0,1,0}{0.03\pm0.01}$ \\
		& GRNN & $33.14\pm1.40$ & $35.66\pm1.53$ & $34.34\pm2.20$ & $\textcolor[rgb]{1,0,0}{41.63\pm2.80}$ & $46.00\pm3.26$ & $37.96\pm1.93$ & $\textcolor[rgb]{0,0,1}{38.88\pm2.63}$ & $32.97\pm2.16$ & $\textcolor[rgb]{0,1,0}{0.03\pm0.01}$ \\ \hline
	\end{tabular}
	\label{Tab:VTFPrediction}
\end{table*}
\begin{figure}[t]
	\centering
	\includegraphics[width=\linewidth]{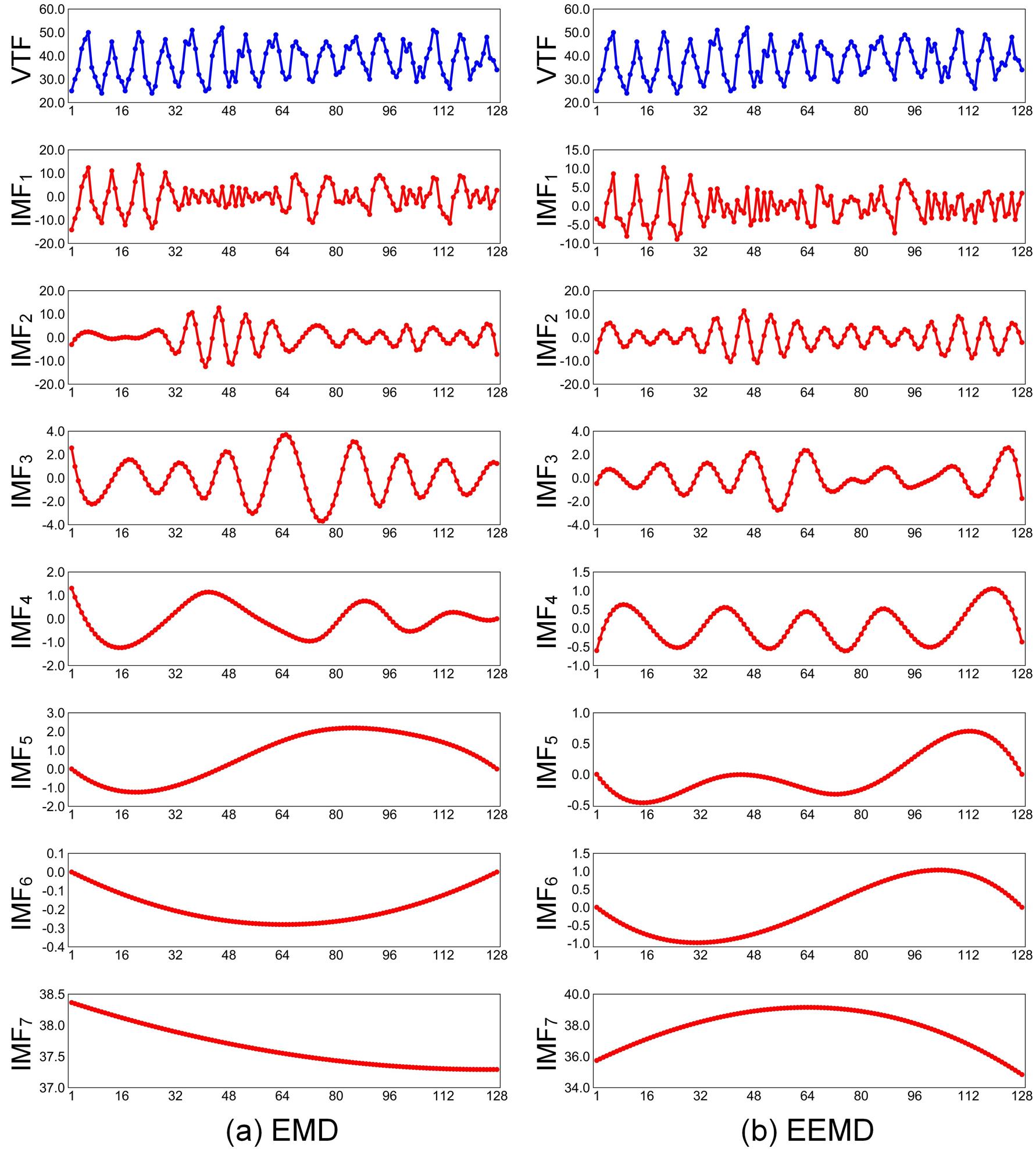}
	\caption{From left to right: (a) EMD- and (b) EEMD-based decompositions of original VTF time series (the $1$st row). It is difficult to predict the first $3$ IMFs with high frequencies due to their properties of weak mathematical regularity. In contrast, the predictions of last $4$ IMFs can be easily obtained due to the significant mathematical regularities.}
	\label{Fig:VTFDecomposition}
\end{figure}
\subsubsection{Prediction Results and Discussion}
It is obvious that ENN produces the highest prediction accuracy and robustness among all NN methods, demonstrated in Figs. \ref{FigurePred2016}-\ref{FigureVarianceRE}. Therefore, our three-step calculation framework is compared with other competing methods (i.e., ENN and EMD+ENN) on PCT prediction from $2017$ to $2026$ in this subsection. The prediction results are displayed in Fig. \ref{FigurePreResult}, which illustrates the comparisons between different prediction frameworks. It can be easily found that our EMD+DTW+ENN has the capacity of robustly implementing the prediction of PCT time series. The original data in Fig. \ref{Figureoridata} illustrate that the trends of PCT in most ports show steady increases from $1996$ to $2016$. From a theoretical point of view, the predicted results from $2017$ to $2026$ should also grow with a steady increase. Fig. \ref{FigurePreResult} shows that our proposed framework generates the predicted PCT time series which are more similar to the theoretical results. Both ENN and EMD+ENN easily suffer from unstable calculation leading to low-quality prediction performance. By taking full advantages of property decomposition and similarity grouping, our proposed three-step calculation method is able to adequately grasp the changing tendency of PCT time series and enhance the prediction accuracy. In particular, EMD decomposes original PCT time series into a sum of finite number of components. Each component becomes less volatile and thus improves the NN-based prediction accuracy. Experiments have illustrated that our proposed method could accurately and robustly predict PCT time series in practice.
\begin{figure*}[t]
	\centering
	\includegraphics[width=\linewidth]{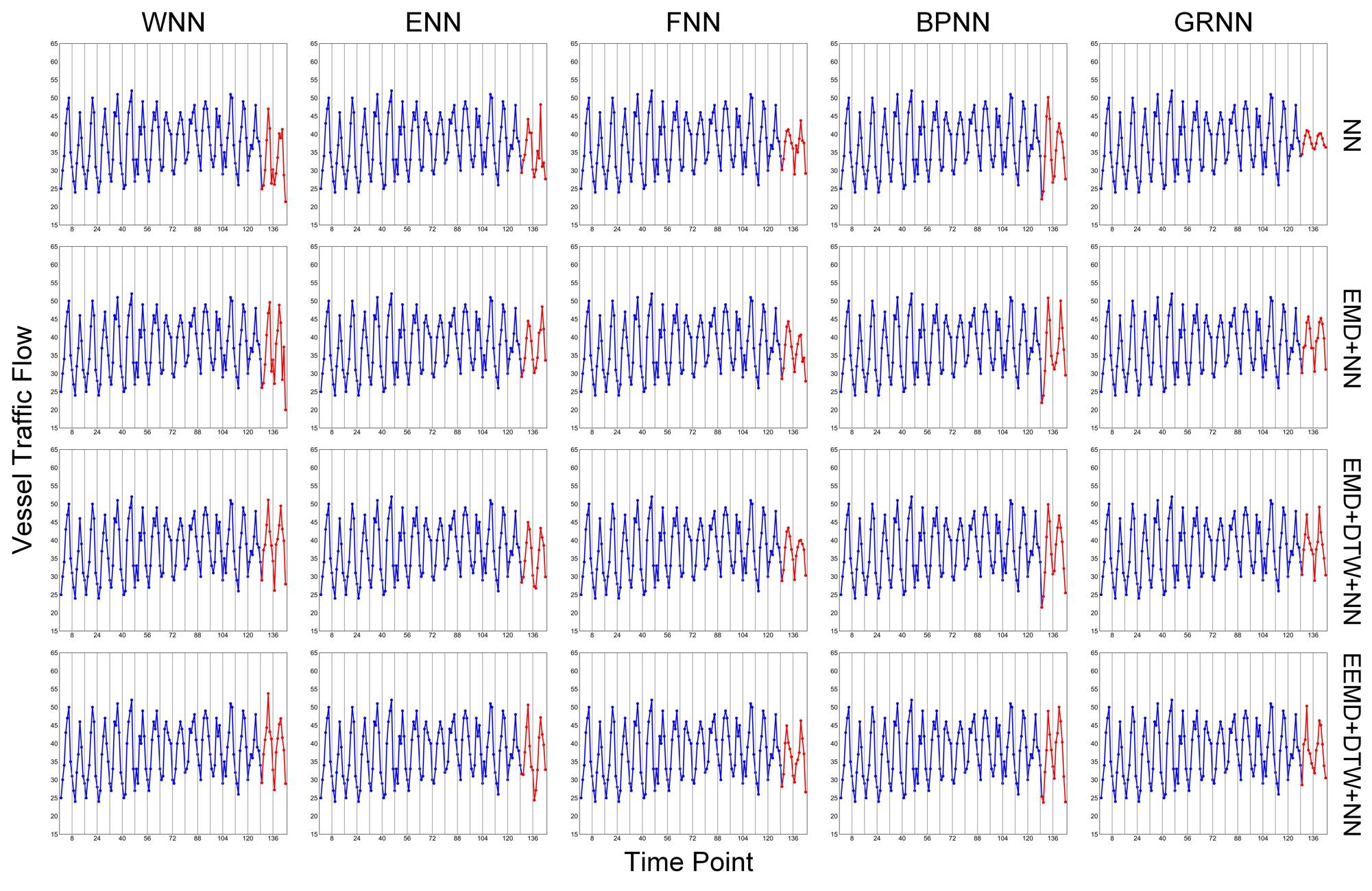}
	\caption{Prediction results of $4$ different frameworks (i.e., NN, EMD+NN, EMD+DTW+NN and EEMD+DTW+NN) on the $17$th and $18$th days (i.e., \textit{Time Point from $129$ to $144$ in the horizontal axis}). Our proposed frameworks have the capacity of adequately grasping the changing tendency of VTF and guaranteeing prediction accuracy.}
	\label{Fig:VTFPred}
\end{figure*}
\subsection{Experiments on Vessel Traffic Flow Prediction}
\label{sec:VTFprediction}
Current research findings \cite{LiuITSC2017,XiaoIEEEITS2017} show that the growth of VTF time series has the predominantly complex, nonlinear and nonstationary properties. The high-quality prediction results are of importance in the fields of maritime management, traffic safety control, and maritime industrial planning, etc. The original VTF data were extracted from the dynamic AIS information at the Wuhan Yangtze Great Bridge. In the case of quantitative evaluation, the VTF time series of $15$ days with an interval of $3$ hours are adopted as the training datasets to optimize the traditional NN models. The VTF data on the $16$th day are considered as the testing datasets to assess the prediction results. Our proposed three-step frameworks EMD+DTW+NN and EEMD+DTW+NN will also be compared with conventional prediction methods NN and EMD+NN. It is well known that the number of IMFs plays a crucial role in decomposition-based prediction performance. As illustrated in Fig. \ref{Fig:VTFDecomposition}, the original VTF data are decomposed into $7$ IMFs to achieve the balance between complexity and performance. Without loss of generality, we observe that the empirical selection of parameters $P = 3$ and $Q = 4$ in Eq. (\ref{Eq:predictionresults}) is able to guarantee high-accuracy prediction.

Table \ref{Tab:VTFPrediction} depicts the quantitative results from our proposed three-step frameworks and other competing prediction methods. It can be observed that NN generates the lowest-accuracy prediction results under almost all conditions. By taking advantage of the characteristics of EMD, EMD+NN significantly promotes traditional NN methods and enhances the prediction accuracy. Furthermore, the DTW-based similarity grouping incorporated into our proposed framework is able to cluster the (time series) segments with high similarities. The grouped segments perform usefully in training the popular NN methods and generating satisfactory prediction results. In Table \ref{Tab:VTFPrediction}, our proposed methods provide the most accurate and robust prediction results. It happens due to the fact that EMD- or EEMD-based property decomposition and DTW-based similarity grouping benefit for predicting non-stationary VTF time series. More detailedly, EEMD+DTW+NN performs better than EMD+DTW+NN since EEMD is able to yield more robust and stable decomposition results. The good performance of our proposed prediction frameworks is further confirmed by the mean and standard deviation values of RE, shown in Table \ref{Tab:VTFPrediction}. Our proposed prediction methods and other competing methods are further adopted to forecast the VTF data on the $17$th and $18$th days. The prediction results are detailedly displayed in Fig. \ref{Fig:VTFPred}, which illustrates the comparisons between different prediction frameworks. It can be observed that both NN and EMD+NN easily suffer from some perturbations leading to prediction quality degradation. In contrast, our proposed methods are able to robustly implement the prediction of VTF time series. This is because that EMD and EEMD can decompose non-stationary time series into high- and low-frequency components to reduce the uncertainties in original data. The self-similarities within high-frequency components are then fully used to enhance the NN-based prediction quality. Experimental results have demonstrated that our proposed framework performs well in non-stationary time series prediction in maritime industry.
\section{Conclusions}
\label{section4}
The reliable prediction of non-stationary time series is of fundamental importance in maritime industries, e.g., economic investment, transportation planning, port planning and design, etc. In this paper, we proposed to develop two three-step calculation frameworks, called EMD+DTW+NN and EEMD+DTW+NN, to guarantee the prediction accuracy and robustness for both port cargo throughput and vessel traffic flow. In particular, the original non-stationary time series were first decomposed into high- and low-frequency components using the EMD and EEMD methods. To take advantage of self-similarities within high-frequency components, these components were further divided into several small (overlapping) segments. The grouped segments with high DTW similarities were selected to form the training dataset for widely-used NN methods to enhance the prediction accuracy of high-frequency components. In contrast, the low-frequency components could be accurately predicted directly using traditional NN methods. The final prediction results were consequently obtained by combining the predicted high- and low-frequency components. Experiments on both port cargo throughput and vessel traffic flow have illustrated the superior performance of our proposed frameworks in terms of prediction accuracy and robustness. Although this work only focuses on predicting non-stationary time series in maritime industry. There is a huge potential to adopt the proposed three-step prediction frameworks to generate satisfactory performance in other industries.
%
%

%
%
%
%
\ifCLASSOPTIONcaptionsoff
  \newpage
\fi

\end{document}